# Sub-diffraction optical manipulation of the chargestate of nitrogen vacancy center in diamond

Xiangdong Chen[1,2,*], Changling Zou[1,2,*], Zhaojun Gong[1,2], Chunhua Dong[1,2], Guangcan Guo[1,2] and Fangwen Sun[1,2]

[1]*Key Lab of Quantum Information, University of Science and Technology of China, Hefei 230026, China*

[2]*Synergetic Innovation Center of Quantum Information & Quantum Physics, University of Science and Technology of China, Hefei, 230026,China*

\* These authors contributed equally.

Correspondence: FW Sun, Key Lab of Quantum Information, University of Science and Technology of China, Hefei 230026, China

E-mail: fwsun@ustc.edu.cn

**As a potential candidate for quantum computation and metrology, the nitrogen vacancy (NV)center in diamond presented both challenges and opportunities resulted from charge state conversion. By utilizing different lasers for the photon-induced charge state conversion, we achieved the sub-diffraction charge state manipulation. The charge state depletion (CSD) microscopy resolution was improved to 4.1 nm by optimizing the laser pulse sequences. Subsequently, the electron spin state dynamics of adjacent NV centers were selectively detected via the CSD. The experimental results demonstrated that the CSD can improve the spatial resolution of the measurement of NV centers for nanoscale sensing and quantum information.**

**Keywords**: Super-resolution microscopy; NV center; Charge state; Photon ionization

# INTRODUCTION

Owing to the stable fluorescence and long coherence time of its spin state, the negatively charged nitrogen vacancy center (NV⁻) in diamond has been studied extensively over past decade. This defect has the potential to be used for quantumcomputation,[1–4] nanoscale metrology[5–8] and biologicalimaging.[9–11] To further extend the study of interaction between multi-NV-center and the nanoscale sensing with NV center, it is necessary to detect and control the NV center spin state dynamics with high spatial resolution.[7, 12–14] Therefore, many optical super-resolution microscopy techniques have been developed to detect single NV center.[13–17] Among these methods, stimulated emission depletion (STED) microscopy[12, 18–20] is one of the most promising. It utilizes a doughnut-shaped laser to produce the position dependent stimulated emission, which changes the fluorescence signal. With STED, the electron spin resonance signals of NV centers has been detected with a resolution lower than the diffraction limit.[12,21,22]

Meanwhile, a new super-resolution microscopy technique was recently developed by Han etal.[15] The authors replaced the stimulated excitation of STED with the dark state pumping of NV centers. The dark state was proven to be the neutral charge NV center ($NV^0$) later by other groups.[23-25] The charge state conversion results in a change of local field in diamond[26] and the spectral diffusion of NV center.[24,27] For high fidelity quantum manipulation, the charge state should be well controlled.[28] Based on the mechanism of charge state conversion and super-resolution microscopy,[25,29,30] we demonstrated the optical manipulation of the charge state of NV center with sub-diffraction resolution. By changing the duration and power of laser pulses, we optimized the charge state depletion (CSD) microscopy to improve the resolution to 4.1 nm without oil immersion lens. With the charge state manipulation, we were able to detect the electron spin state dynamics of NV centers with sub-diffraction resolution, which can be used for the nanoscale sensing of electromagnetic field and biological molecules. Because charge state conversion changes the local field in diamond,[2,26] CSD microscopy method can potentially be used for the study of spin state quantum coherent dynamics and of the interaction between coupled spin systems in diamond.[2,13,31]

# MATERIALS AND METHODS

The NV center in diamond consists of a substitutional nitrogen atom and an adjacent vacancy. Two charge states are usually observed: $NV^0$ and NV⁻ (Figure 1a). Both NV⁻ and $NV^0$ are stable in the absence of optical excitation.[15,32] The electron spin of NV⁻ ground state can be optically initialized and detected.[1] The coherence time of spin state could exceed several ms.[33] However, the $NV^0$ charge state does not show such properties. As shown in Figure 1c, the photons of $NV^0$ and NV⁻ are different in wavelength. Therefore, we are able to distinguish the two charge states by measuring the NV center fluorescence. The conversion between these two charge states can be seen as the fluorescence on and off processes, which can be used for reversible saturable optical fluorescence transitions microscopy.

The samples used in this experiment were ⟨100⟩ surface diamond plates with nitrogen concentration lower than 5 ppb. The NV centers were produced by $N^+$ ions implantation and subsequently annealing. The optical system used to locate and detect the NV center is shown in Figure 1b. Laser pulses with various wavelengths and beam shapes were switched using acoustic optical modulators, then focused on the sample through an objective (N.A. = 0.9). The fluorescence emission of $NV^0$ and $NV^-$ was split using a long pass dichroic mirror, and was further filtered using a short pass filter and a long pass filter, respectively.

## RESULTS AND DISCUSSION

### Optical manipulation of the charge state

Various methods have been developed to control the charge states of NV centers,[25,27,30,34-37] with the most convenient method being high speed optical manipulation. To demonstrate photon-induced charge state conversion between $NV^0$ and $NV^-$, the NV center ensemble fluorescence emission of both $NV^0$ and $NV^-$ charge states was detected. As shown in Figure 1d, the fluorescence intensities of the two charge states changed in opposite manners when the NV center was pumped with a 0.68 mW 532 nm laser. The fluorescence changing rates of the two charge states were almost the same (about 2.7 $\mu s^{-1}$ in Figure 1d). In contrast, the excited state lifetimes of $NV^-$ and $NV^0$ are both around 13 ns in our experiments. As the fluorescence intensity is directly proportion to the population of charge state, the fluorescence changing rate can be simply presented as the charge state conversion rate. This result indicates that the charge state can be switched from $NV^0$ to $NV^-$ using a 532 nm laser.

Usually, both the ionization($NV^-$ to $NV^0$) and recharging ($NV^0$ to $NV^-$) processes are able to be excited using the same laser. The steady state population is determined by both the ionization and recharging rates, which are significantly changed by the wavelength and power of laser. We measured the charge state population by the single shot charge state readout method.[25,30] In our experiment, the highest population of $NV^-$ (about 75%) was obtained by excitation with a 532 nm laser, while the lowest population of $NV^-$ (about 5%) was obtained using 637 nm laser (Figure 1e).Therefore, these two lasers were chosen to initialize and change the charge state population of the NV centers. The fidelity of charge state initialization would be higher than that in Han's work,[15] where 473/592 nm laser was used to initialize the NV center charge state. Previous experiments demonstrated that the charge state conversion is mainly a two-photon process with laser wavelength longer than 500 nm.[25,30,36] The NV is first pumped to the excited state of $NV^-$ or $NV^0$ , then an electron is released to the conduction band during ionization or captured from the valance band during recharging.[24] The nonlinear power dependence of the conversion rates in Figure1f was obtained by repeating the measurement of Figure 1d with different laser power. The results prove that the charge state conversions with 532 and 637 nm are two-photon processes (see supplementary material for details).[23,25,30]

Simply, we can write the rate equations during charge state conversion as

$$\dot{\rho}_0 = -\gamma_r \rho_0 + \gamma_i \rho_-, \qquad (1)$$

$$\dot{\rho}_- = +\gamma_r \rho_0 - \gamma_i \rho_-, \qquad (2)$$

where $\rho_0$ and $\rho_-$ are the population of NV$^0$ and NV$^-$, respectively. $\gamma_r$ and $\gamma_i$ are the recharging and ionization conversion rates, respectively. If only the photons of NV$^-$ are detected, the fluorescence intensity can be written as:

$$\eta \propto \rho_- = \rho_{-,st} + (\rho_{-,in} - \rho_{-,st})e^{-\gamma\tau}, \qquad (3)$$

where $\gamma = \gamma_r + \gamma_i$ is the experimentally measured fluorescence changing rate, as in Figure 1d. $\rho_{-,st} = \gamma_r/(\gamma_r + \gamma_i)$ is the steady state NV$^-$ population with laser pumping, and $\rho_{-,in}$ is the initial NV$^-$ population. $\tau$ is the laser duration. Therefore, the high spatial resolution charge state manipulation requires a position-dependent charge state conversion rate $\gamma(r)$ or a position-dependent steady state population $\rho_{-,st}(r)$.

According to the experimental results in Figure 1e, the steady state populations change little with the power of 532 nm and 637 nm lasers. In contrast, the charge state conversion rates increase with the power of the laser (Figure 1 f). Hence, a doughnut-shaped laser beam is used to produce the position-dependent charge state conversion rate. The power intensity at the beam center approaches zero. We consider two NV centers, one at the beam center with position $r_0$, and the other at $r_0 + \Delta r$. Because $\gamma(r_0) = 0$, the fluorescence difference between the two NV centers is $\left|\eta_{r_0} - \eta_{r_0 + \Delta r}\right| \propto \left|\rho_{-,in} - \rho_{-,st}\right| \cdot \left(1 - e^{-\gamma(r_0 + \Delta r)\tau}\right)$. Therefore, for any $\Delta r$, the fluorescence difference $\left|\eta_{r_0} - \eta_{r_0 + \Delta r}\right|$ is saturable, and can be increased with power and duration of laser.

**CSD microscopy**

Next, we will outline the procedure for sub-diffraction CSD microscopy imaging. Three different laser pulses were used in the experiment, as shown in Figure 2a and b. First, a Gaussian beam (G) laser was used to initialize the charge state in the focus region. Second, a doughnut-shaped (D) laser pulse was applied. The wavelength of the D laser was different from that of the G laser. This laser was able to deplete the $NV^0$ or $NV^-$ charge state except for the NV at the beam center. Finally, the charge state was detected with a 0.1 mW 589 nm laser. The duration of 589 nm laser was 5 μs. The microscopy image was obtained by detecting the fluorescence of $NV^-$ and scanning the sample with a piezo stage. All of the lasers were circularly polarized to obtain the same conversion rates for the NV centers with different symmetry axes.

Specifically, for high-resolution ionization manipulation(change NV center to a high $NV^0$ population state), we used a 637 nm G laser and a 532 nm D laser (Figure 2a). The 637nm laser initialized the NV with an $NV^0$ population of approximately 95%. The 532 nm D laser converted $NV^0$ to $NV^-$ with position-dependent conversion rate. Example of microscopy image is shown in Figure 2a. NV centers are presented as dark spots in the image. In contrast, using a 637 nm D laser and a532 nm G laser, we achieved the higher solution recharging manipulation (change NV center to a high $NV^-$ population state)(Figure 2b). The NV center would be initialized to $NV^-$ with probability about 75%, and converted to $NV^0$ with position dependent rate. NV centers are shown as bright points in the fluorescence microscope image. To distinguish the imaging using different laser sequences, the CSD microscopy using a 637 nm D laser can be termed as recharging-CSD (rCSD), and the 532nm D laser CSD is termed as ionization-CSD (iCSD).

The effective point-spread-function (PSF) of the CSD image is represented by $h(r) = h_{\text{det}}(r)\rho_-(r)$, as $h_{\text{det}}(r)$ is the detection PSF. For the two-photon charge state conversion, the charge state conversion rate should be quadratically dependent on the power of laser, we simply write the conversion rate as $\gamma \approx \alpha I^2$ (see details in supplementary material), where $I$ is the power of laser. The most simple function to depict the detection PSF and the doughnut conversion laser intensity is the standing wavefunction:[38,39]

$$h_{\text{det}}(r) = C\cos^2(\pi r/\omega_{\text{det}}), \tag{4}$$

$$I_D(r) = I_{\max}\sin^2(\pi r/\omega_D). \tag{5}$$

For simplicity, we assume $\omega_{\text{det}} \approx \omega_D$ (the width of detection Gaussian beam laser approximates the width of D laser). The resolution of CSD can be approximated by a Taylor series:

$$\Delta r = \frac{2\omega_D}{\pi}\sqrt{\frac{-3+\sqrt{3}\sqrt{6\beta+1}}{2(3\beta-1)}}, \tag{6}$$

where $\beta = \alpha I_{\max}^2 \tau$. It indicates that increasing the power and duration of D laser could improve CSD resolution, as shown in Figure 2c-d. And the microscopy resolution is theoretically unlimited. For the same power, the charge state conversion rate of the 532 nm laser is higher than that of the 637 nm laser, as shown in Figure 1f, which might be resulted from the wavelength dependent absorption cross section of NV center at room temperature.[25,36] Therefore, the resolution of iCSD is better than rCSD for the same power and duration of the D laser.

In reality, the improvement of CSD resolution is still limited by the power of laser and stability of piezo stage. Figure 2e illustrates the best resolution in our experiment. The insert shows the iCSD image with the FWHMs in two directions about 4.1 nm and 7.8 nm, respectively. As the drift of piezo stage is estimated to be lower than 0.15nm/s, the affection of drift to the resolution is around 1 nm. The difference between two axes might be caused by the inhomogeneous intensity distribution of the doughnut laser beam center or the drift of stage.[22] To further improve the spatial resolution, a high numerical aperture oil objective could be used in the future. Such an objective would produce a higher charge state conversion rate gradient than dry objective.

Same with the STED method, CSD can resolve the location of high density NV centers better than confocal microscopy(Figure 3a-b). The lasers for charge state conversion of CSD microscopy are much weaker than the lasers in STED microscopy with NV.[19,21] Furthermore, our experiments showed that an optimized G laser duration would also improve the spatial resolution without changing the D laser duration(see supplementary material for details). Therefore, the durations of both the G laser and the D laser should be carefully adjusted to improve the imaging quality.

**Spin state detection**

We have demonstrated that the NV center charge state can be controlled using CSD microscopy. It enables us to detect the spin state of NV with sub-diffraction spatial resolution. The energy levels of the NV$^-$ triplet ground state, as shown in Figure3c, can be changed by the magnetic field and temperature. The two resonant frequencies of the spin state transition can be detected using the optically detected magnetic resonance (ODMR)method. To compare the difference between CSD ODMR and confocal ODMR, we detected the ODMR signals of two adjacent NV centers with different symmetry axes, indicated as NVA and NV B in Figure 3b. The pulse sequences are shown in Figure 3d. The 532 nm Gaussian beam laser was used for both initialization and detection NV center. Using 532 nm laser for detection will not affect the spatial resolution of rCSD ODMR, as the ODMR signal's contrast between different NV centers will only be determined by the charge state population before the microwave is applied. The resonant signals with different magnetic field were measured (Figure 3e-f). The confocal ODMR in the focus area did not distinguish the electron spin resonance signals of the different NV centers. In contrast, using rCSD

ODMR, the resonant frequencies of each NV were separately determined. Only the NV centers at the beam center of D laser will remain to be NV⁻ and contribute to the resonant signals of rCSD ODMR. In Figure 3e and f, the noise of rCSD ODMR seems to be higher than confocal ODMR with the same acquisition time. This is caused by the long duration of D laser, which compresses the time for fluorescence detection. Though we did not detect iCSD ODMR, we expected this method could also separate the different NV centers signals. The iCSD ODMR would need one more laser beam than rCSD ODMR, as the 637 nm Gaussian beam is not suitable to be used for detecting the fluorescence of NV⁻. Here, the magnetic field was supplied by a permanent magnet with a low magnetic field gradient. Supposing the magnetic field at NV A was the same as that at NV B, we calculated the magnetic field vector in Figure 3e,[40] as the magnitude $|B_1| \approx 25.5G$, and direction $\theta_{1,A} \approx 73°$, $\theta_{1,B} \approx 60°$. Here, $\theta$ shows an angle between the magnetic field and the symmetry axis of the NV center. Another magnetic field in Figure 3f was $|B_2| \approx 48.8G$, $\theta_{2,A} \approx 76°$, $\theta_{2,B} \approx 42°$. The CSD ODMR results revealed the change in both magnitude and direction of magnetic field.

Although STED ODMR is also able to detect NV centers resonant frequencies with sub-diffraction resolution,[12] the principles of STED ODMR and CSD ODMR are different. In the STED method, the resonant microwave pulses affect all of the NV centers, but only one NV center's fluorescence is detected. Unlike the STED ODMR, the CSD ODMR high-resolution resonant signal detection relies on charge state manipulation. If an NV center is at NV⁰ charge state, no ODMR signal will be detected. The difference between STED ODMR and CSD ODMR might emerge in the results of coupled NV centers. For NV A and NV B in Figure 3, the resonant frequencies are different. The spin states of these two NV⁻ centers can be separately controlled using different frequencies microwave pulses even without the help of CSD microscopy. However, NV C and NV D in Figure 3b showed the same resonant frequency(Figure4a). To selectively detect the electron spin state dynamics of each NV center, we used the CSD method. The Rabi oscillation of an individual NV center was measured using rCSD method, as shown in Figure 4b. The individual Ramsey fringe signals were also detected to obtain spin state coherence information (Figure4c). Both of the NV centers showed short spin coherence times. It might result from the local implantation damage.[41] As the distance between these two NV centers was larger than 100 nm, the affection of dipole interaction was not observed. The results demonstrate that the spin state information of NV can be detected with sub-diffraction spatial resolution using CSD microscopy.

The charge state conversion will destroy the information of NV spin state. However, recent experiments demonstrated that the charge state conversion of NV centers could induce a detectable change in the local field in diamond,[2,26] which might affect the dynamics of the spin system that is coupled to the NV center. Subsequently, we expect that CSD could enable us to study the NV center spin state dynamics by changing the charge state of adjacent NV centers in future work. We can use different D lasers to change the adjacent NV to NV$^-$ or NV$^0$ , and keep the charge state of NV at beam center to be NV$^-$ . This proposal requires us to maintain the spin state of NV at beam center while changing the charge state of other NVs. However, for the lasers used in this experiment, the charge state conversion is usually much slower than the NV$^-$ spin state polarization.[30] The spin state of NV$^-$ might be polarized to $m_s = 0$ before we can change the charge state of an adjacent NV center. To solve this problem, one method would be to combine CSD microscopy with spin-RESOLFT microscopy.[14] Another method would be to use a short wavelength laser (e.g., 405 nm) as the D laser. Because the NV center will not be pumped to the excited state of NV$^-$ during a single-photon charge state conversion process,[25,30] the spin state of NV$^-$ would not be changed before the charge state is switched.

## CONCLUSIONS

In summary, we demonstrated sub-diffraction charge state manipulation of the NV centers in diamond. Laser beams with different wavelengths and shapes were used to control and detect the charge state. The best spatial resolution achieved in our experiment is approximately 4.1 nm, which can be further improved. Using CSD microscopy, we measured the resonance signal and coherent dynamics of NV center spin state with sub-diffraction spatial resolution, which could be used for nanoscale sensing. In future, we expect that high-resolution NV charge state manipulation could help to control the spin state dynamics of NVs and to switch interactions between NV centers.

### ACKNOWLEDGEMENTS

This study was financially supported by National Basic Research Program of China (GrantNo.2011CB921200), the Knowledge Innovation Project of Chinese Academy of Sciences (Grant No.60921091), the National Natural Science Foundation of China (Grant No.11374290), the Program for New Century Excellent Talents in University, the Fundamental Research Funds for the Central Universities, and the Foundation for the Author of National Excellent Doctoral Dissertation of China.

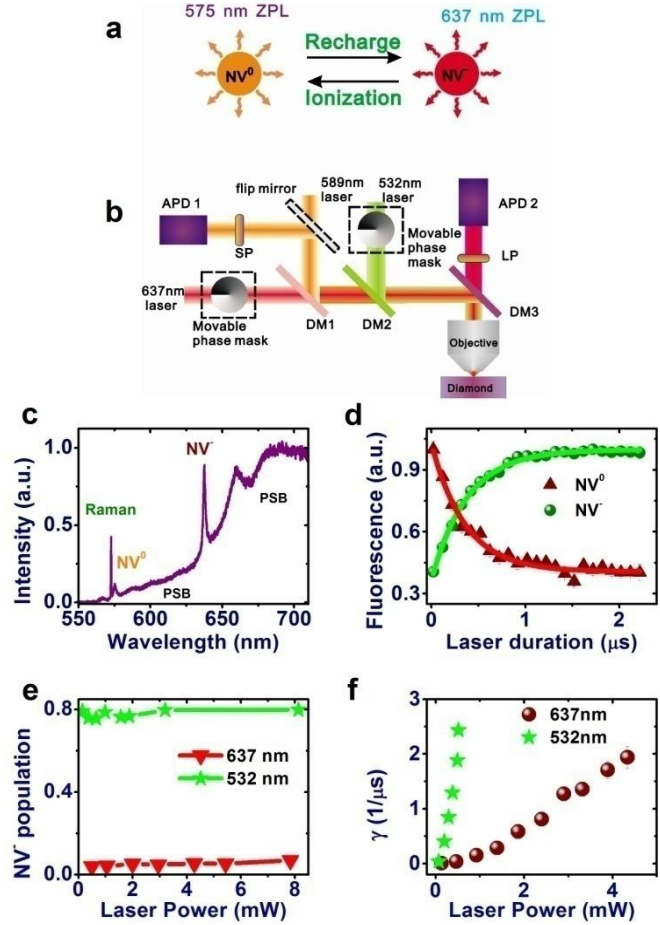

**Figure 1 (a)** A sketch depicted the charge state conversion between NV$^-$ and NV$^0$. **(b)** Schematics of the experimental configuration. DM: long pass dichroic mirrors; APD: avalanche photodiode. The lasers and fluorescence emission were combined and split using DM1 (edge wavelength 605 nm), DM2(edge wavelength 536.8 nm) and DM3(edge wavelength 658.8 nm). The fluorescence of NV$^-$ was detected by APD 2 with a long pass filter (LP, edge wavelength 668.9 nm), and the fluorescence of NV$^0$ was detected by APD 1 with a short pass filter (SP, edge wavelength 668.9 nm). Two phase masks were used to produce the doughnut-shaped laser beams. The 637 nm and 532 nm lasers were used to initialize or switch the charge state. And the 589 nm laser is for the detection of the charge state. **(c)** Fluorescence spectrum of the NV center ensemble excited by 532 nm laser at temperature about 250 K. The charge states of NV are marked by different ZPLs(zero phonon lines). Phonon sideband (PSB) of each charge state presents photons at longer wavelength. **(d)** Charge state conversion process of NV ensemble pumped by a 532 nm laser. The charge states were initialized using a 637 nm laser. **(e)(f)** Power dependence of the steady charge state population and charge state conversion rates ($\gamma = \gamma_r + \gamma_i$). All of the lasers were Gaussian beam laser. The charge state conversion rates were obtained by fitting the results of NV$^-$ fluorescence in (d) with Eq.3.

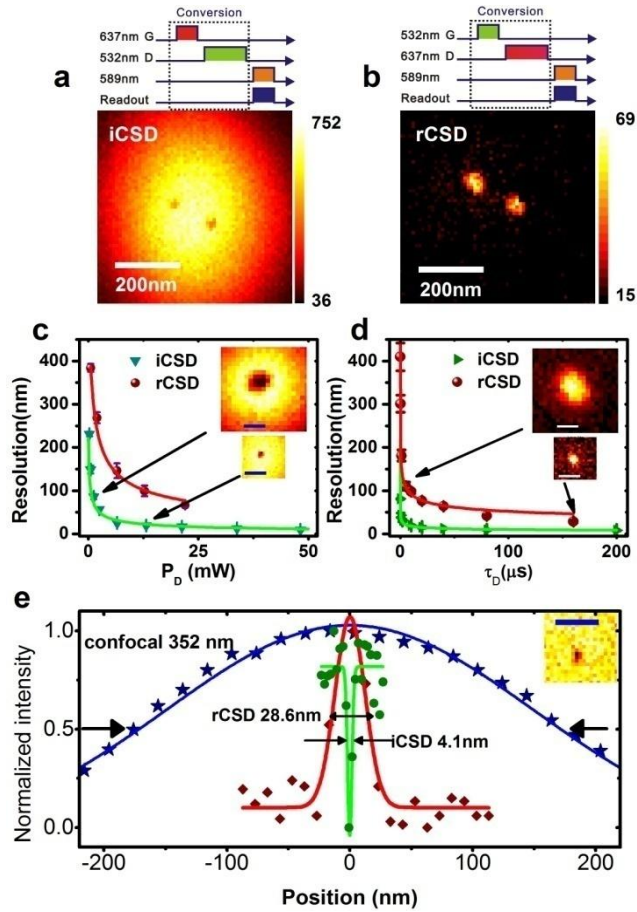

**Figure 2 (a)(b)** Laser sequences and example images for the high resolution charge state ionization and recharging manipulation, respectively. **(c)(d)** The resolution (presented by FWHM) improved by increasing the D laser power or duration. The solid lines are the fits with Eq. 6 . The duration of the D lasers in **(c)** is 40μs. The D laser in **(d)** is 34 mW the 532 nm laser for iCSD, and 22 mW 637 nm laser for rCSD. The inserts in **(c)(d)** are the images of iCSD with different power and rCSD with different duration, respectively. The scale bar is 100 nm. **(e)** The profiles of confocal (blue star dots) and CSD images. In our experiment, the best rCSD resolution was approximately 28.6 nm (22 mW 637 nm doughnut laser with 160μs duration), shown as red square dots. And the best iCSD resolution was 4.1nm (48 mW 532 doughnut laser with 200 μs resolution), shown as the green circular dots. The insert shows the iCSD image of the best resolution. The scale bar is 30 nm.

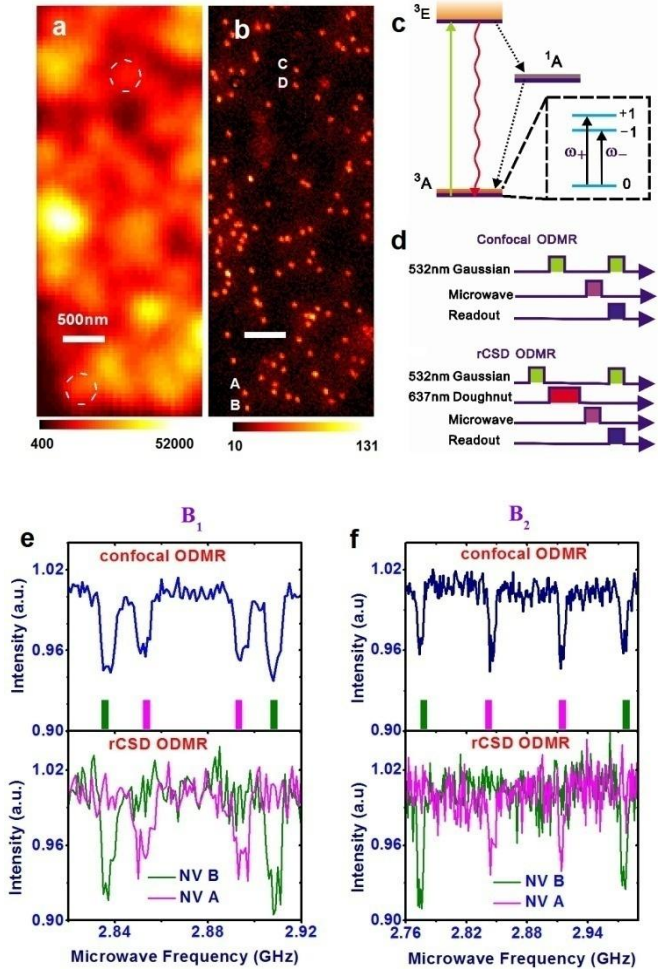

**Figure 3 (a)(b)** Confocal microscopy and rCSD microscopy images of the same area in the diamond sample. The confocal microscopy was obtained by pumping and detecting NV with 532 nm Gaussian beam laser. The rCSD microscopy was obtained with method shown in Figure 2b. The D laser was 22 mw 637 nm laser with 50 μs duration. The dwell time of every scanning step was 50 ms. **(c)** The energy levels structure of NV$^-$. The insert shows the details of the triplet ground state. **(d)** The pulse sequence used to detect the ODMR signals. The signal of ODMR was detected by 0.7 mW 532 nm laser with 300 ns duration. **(e)(f)** The ODMR signals of two NV centers. The resonant frequencies are marked with pink (NV A) and green (NVB) bars. With magnetic field in **(e)**, the resonant frequencies were $\omega_{+,A} = 2.894 GHz$, $\omega_{-,A} = 2.852 GHz$, $\omega_{+,B} = 2.908 GHz$, $\omega_{-,B} = 2.837 GHz$. In **(f)**, the resonant frequencies were $\omega_{+,A} = 2.914 GHz$, $\omega_{-,A} = 2.845 GHz$, $\omega_{+,B} = 2.975 GHz$, $\omega_{-,B} = 2.774 GHz$.

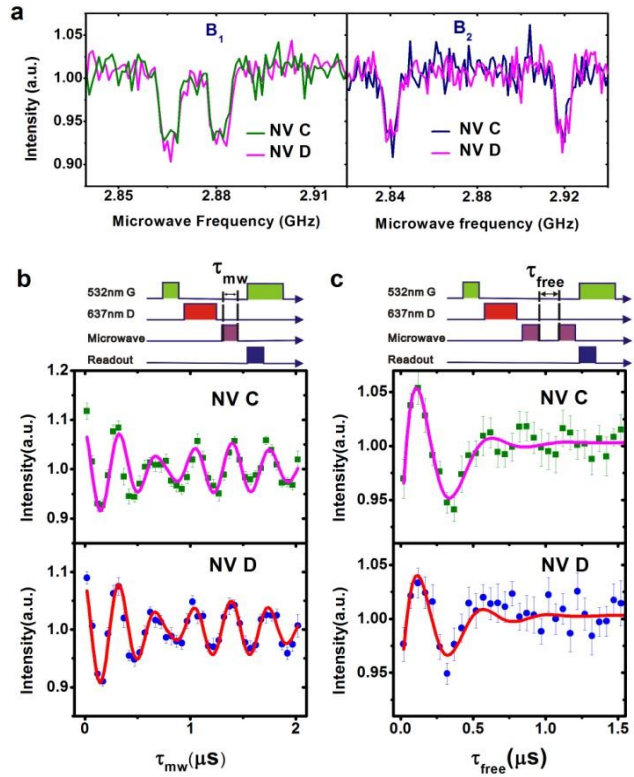

**Figure4 (a)** rCSD ODMR signals of NV C and NV D in Fig. 3 . The different ODMR results for different magnetic fields revealed that the two NV centers had the same symmetry axis. **(b)** The electron spin state Rabi oscillation using rCSD. **(c)** The Ramsey fringes of the NV⁻ spin states using rCSD method.

# Supplementary material

# Sub-diffraction optical manipulation of the chargestate of nitrogen vacancy center in diamond

Xiangdong Chen[1,2,*], Changling Zou[1,2,*], Zhaojun Gong[1,2], Chunhua Dong[1,2], Guangcan Guo[1,2] and Fangwen Sun[1,2]

[1]*Key Lab of Quantum Information, University of Science and Technology of China, Hefei 230026, China*

[2]*Synergetic Innovation Center of Quantum Information & Quantum Physics, University of Science and Technology of China, Hefei, 230026,China*

**S1. Method**

The experimental measurements were performed with a home built confocal system with different wavelengths lasers (532 nm, Compass 315m, Coherent; 589 nm, MGL-III-589nm, New Industries Optoelectronics; 637 nm, MRL-III-637nm, New Industries Optoelectronics). Laser pulses were switched by acoustic optical modulators (MT200-0.5-VIS, AA). The doughnut laser beam was produced by vortex phase mask (VPP-1a, RPC photonics). The phase masks will be removed if the Gaussian-shaped laser is needed. Three-dimensional translation stages were used to align the laser beams. The sample was located by a pizeo-stage (P-733.3CD, PI). The fluorescence was collected by an objective (N.A.=0.9, Nikon) and detected by APDs (SPCM-16, PerkinElmer). The fluorescence of NV- was filtered by long pass filter (Semrock LP02 664RU), while the fluorescence of NV0 was filtered by short pass filter (Semrock FF01 612SP). The photon number was counted by data acquisition cards (PCI-6602 and USB-6343, National Instruments).

**S2. The power dependence of CSD resolution**

Charge state conversion for 532 nm and 637 nm is a two-photon process, which has two steps. Ionization: NV- is firstly pumped from ground state to excited state by absorbing one photon; then one electron is excited in to the conduction band of diamond by another photon. Finally, the NV is changed from NV- to NV0. Recharging: NV0 is pumped to excited state first; then a second photon is absorbed, and one electron is obtained from the valence band. Finally, the NV is changed from NV0 to NV-. The charge state conversion rate is quadratically dependent on the power of weak laser. But if the power of laser is too high, the transition between excited state and the ground state would be saturated. Then the conversion rate would be linearly dependent on the laser power [1]. In our experiment, the saturation power of 637 nm laser is about 5-10 mW, while the saturation power of 532 nm laser is about 0.4-0.5 mW.

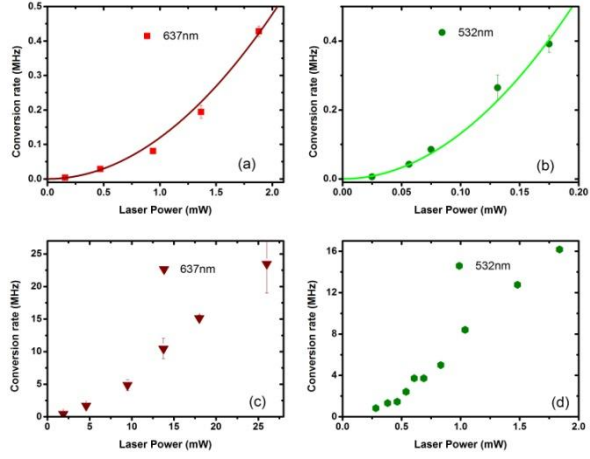

*Figure 1S: The power dependence of charge state conversion rates ($\gamma = \gamma_i + \gamma_r$) for 637 nm and 532 nm laser. (a) (b) are results with weak power. (c) (d) are results with strong power.*

In the experiments of CSD, the durations of D laser were set in range from several μs to hundreds of μs. As shown in Figure 1S, the charge state conversion rate is quadratically dependent on the laser power in this range. Therefore, we can simply describe the power dependence of charge state conversion rate as $\gamma = \alpha \cdot I^2$. Then the charge state of NV center is $\rho_- = \rho_{steady} + (\rho_{initial} - \rho_{steady})e^{-\alpha \cdot I^2 \tau}$. Using standing wave function to depict the beam shapes of detection laser and D laser [2] as in the main text, we obtained the PSF of CSD image:

$$h(r) = C\cos^2\left(\frac{\pi r}{w_D}\right)\rho_{steady} + C\cos^2\left(\frac{\pi r}{w_D}\right)(\rho_{initial} - \rho_{steady})e^{-\alpha \cdot \tau \cdot [I_{max}\sin^2(\frac{\pi r}{w_D})]^2}.$$

The first part can be seen as a fluorescence background with confocal FWHM. For the resolution below confocal microscopy, we neglected the first part. Then the FWHM of CSD can be obtained by solving:

$$\cos^2\left(\frac{\pi r}{w_D}\right)e^{-\alpha \cdot \tau \cdot [I_{max}\sin^2(\frac{\pi r}{w_D})]^2} = \frac{1}{2}.$$

Expand the left side with a Taylor series to fourth order:

$$1 - \left(\frac{\pi}{w_D}\right)^2 r^2 + \left(\frac{1}{3} - \alpha\tau I_{max}^2\right)\left(\frac{\pi}{w_D}\right)^4 r^4 = \frac{1}{2}.$$

Then we get the resolution of CSD:

$$\Delta r = \frac{2\lambda_D}{\pi}\sqrt{\frac{-3+\sqrt{3}\sqrt{6\alpha\tau I_{max}^2+1}}{2(3\alpha\tau I_{max}^2-1)}}.$$

This results shows that the power dependence and duration dependence of resolution are different.

## S3. Comparison between the resolution of iCSD and GSD ( ground state depletion ) based on saturation

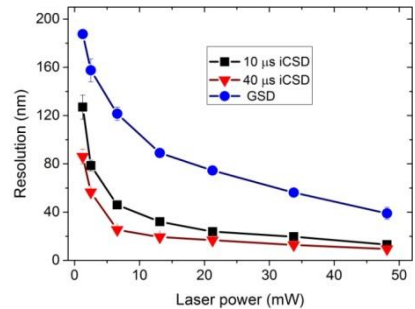

*Figure 2S: The resolution of iCSD and GSD. The durations of D laser in iCSD were 10 $\mu$s or 40 $\mu$s.*

In reference [3], the authors presented a GSD microscopy based on the fluorescence saturation of NV center. In their method, the microscopy image was directly obtained by detecting the NV with a doughnut-shaped laser. The image of GSD is similar with that of iCSD in our results, as NV is presented by a dark point. In the GSD method, the fluorescence of NV is only determined by the power of laser. The saturation power of 532 nm laser is about 0.4 mW in our experiment. And the charge state conversion time of the saturation power is shorter than 1 $\mu$s. Therefore, for D laser duration longer than 1 $\mu$s, the resolution of iCSD microscopy should be better than GSD. We measured the GSD resolution and iCSD resolution at the same time, as shown in Figure 2S. The results show that the resolution of iCSD is much better than GSD in our experiment.

## S4. The charge state manipulation resolution with Gaussian beam laser

For adjacent NV centers, the largest NV- population contrast is determined by the contrast of charge state conversion rate, which is limited by the position dependence of laser intensity. Therefore, the resolution of charge state manipulation with Gaussian beam laser cannot overmatch that with Doughnut laser. In Fig.3S (a), it shows the charge state manipulation with different duration of 532 nm Gaussian laser pulse. The results indicate that the resolution with Gaussian lasers is also changed by the duration of laser. The best resolution (about 240nm) was obtained at 100 ns duration of 0.68 mW 532 nm laser.

Therefore, G laser of CSD microscopy also affect the resolution of CSD. For the rCSD microscopy, the optimized 532 nm G laser (100 ns duration) can improve the resolution of CSD without the changing of D laser, as shown in Fig. 3S (b).

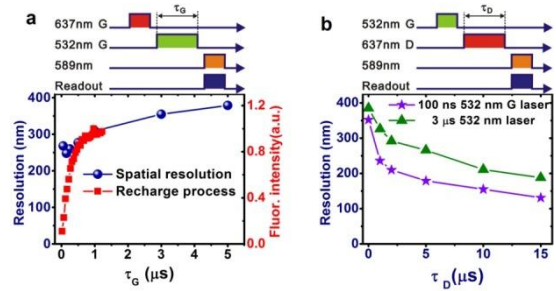

*Figure 3S: (a) The charge state manipulation with 0.68 mW 532 nm Gaussian beam laser. Blue circle points presented the resolution of image. Red square points presented the fluorescence intensity of NV- duration the charge state conversion with 532 nm laser. (b) The resolution CSD microscopy with 532 nm G laser and 637 nm D laser. The power of 532 nm was 0.68 mW, and the 637 nm laser was 4 mW.*

**S5. The polarization dependent charge state conversion rates**

For 532 nm and 637 nm laser, the charge state conversion is attributed to the two-photon process. The laser polarization dependence of the charge state conversion rates was observed in the experiments. For linearly polarized laser, the maximum conversion rate could be more than two times larger than the minimum conversion rate, as shown in Fig.4S. As a result, the CSD resolution is significantly changed by the polarization of laser. In addition, the resolution of CSD by linearly polarized laser is orientation dependent. For NV A and NV B in Fig.4S, the resolutions of rCSD are different as the two NV centers have different symmetry axes.

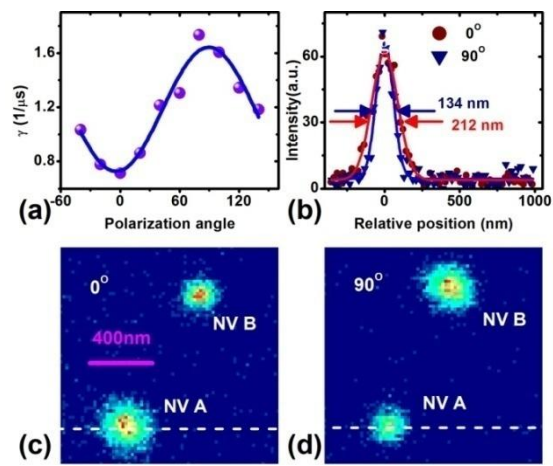

*Figure 4S: (a) The charge state conversion rate of NV A pumped by 4 mW 637 nm Gaussian laser. (b) The fluorescence on the white lines of images. (c)(d) The images of rCSD with different polarization 15 mW 637 nm D laser with duration 2 $\mu s$.*